\documentclass[18pt,a4paper,onecolumn,fleqn,usenatbib,useAMS,referee]{mnras}

\usepackage{newtxtext,newtxmath}

\usepackage[T1]{fontenc}
\usepackage{ae,aecompl}

\DeclareRobustCommand{\VAN}[3]{#2}
\let\VANthebibliography\thebibliography
\def\thebibliography{\DeclareRobustCommand{\VAN}[3]{##3}\VANthebibliography}

\usepackage{graphicx}	
\usepackage{amsmath}	
\usepackage{multicol}   
\usepackage{bm}		    
\usepackage{pdflscape}	
\usepackage{float}
\usepackage{caption}

\usepackage{placeins}
\usepackage{tabularx,booktabs} 

\title[LASER]{Density and infrared band strength of interstellar carbon monoxide (CO) ice analogues} 

\author[C. Gonz\'{a}lez D\'{\i}az et al.]{
C. Gonz\'{a}lez D\'{\i}az,$^{1}$\thanks{E-mail: cgonzalez@cab.inta-csic.es (CGD)}
H. Carrascosa,$^{1}$
G. M. Mu\~{n}oz Caro,$^{1}$
M. \'{A}. Satorre,$^{2}$
and Y.-J. Chen$^{3}$
\\
$^{1}$Centro de Astrobiolog\'{\i}a (CSIC-INTA), Ctra. de Ajalvir, km 4, Torrej\'on de Ardoz, 28850 Madrid, Spain\\
$^{2}$Centro de Tecnolog\'{\i}as F\'{\i}sicas, Universitat Polit\`{e}cnica de Val\`{e}ncia, Plaza Ferr\'{a}ndiz-Carbonell, 03801 Alcoy, Spain\\
$^{3}$Department of Physics, National Central University, Jhongli City, Taoyuan County 32054, Taiwan}

\date{Accepted XXX. Received YYY; in original form ZZZ}

\pubyear{2022}

\begin{document}
\label{firstpage}
\pagerange{\pageref{firstpage}--\pageref{lastpage}}
\maketitle

\begin{abstract}
The motivation to study experimentally CO ice under mimicked interstellar conditions is supported by the large CO gas abundances and ubiquitous presence of CO in icy grain mantles. Upon irradiation in its pure ice form, this highly stable species presents a limited ion and photon-induced chemistry, and an efficient non-thermal desorption. Using infrared spectroscopy, single laser interference, and quadrupole mass spectrometry during CO ice deposition, the CO ice
density was estimated as a function of deposition temperature. Only minor variations in the density were found. The proposed methodology can be used to obtain the density of other ice components at various deposition temperatures provided that this value of the density is known for one of these temperatures, which is typically the temperature corresponding to the crystalline form. The apparent tendency of the CO ice density to decrease at deposition temperatures below 14 K is in
line with recently published colorimetric measurements. This work allowed to revisit the value of the infrared band strength needed for calculation of the CO ice column density in infrared observations, $8.7 \times 10^{-18} ~ {\rm cm ~ molecule}^{-1}$ at 20 K deposition temperature.
\end{abstract}

\begin{keywords}
astrochemistry -- radiation mechanisms: non-thermal -- methods: laboratory: solid state -- techniques: spectroscopic -- ISM: molecules -- infrared: ISM
\end{keywords}


\section{Introduction} \label{sect:intro}

Carbon monoxide (CO) gas abundances in dense clouds suggest that 90 \% of the CO molecules leave the gas phase along the line of sight, and over 99 \% of them should deplete in the core nucleus \citep{Caselli_1999} due to CO freeze out onto dust grains forming ice mantles (e.g., \cite{Pontoppidan_2008}). Protostar formation heats the surrounding environment and triggers the thermal desorption of ice mantles \citep{Cazaux_2003}. 
CO is considered to be one, if not the most, abundant species on the top ice layer covering inter- and circum-stellar dust grains observed toward the coldest regions.
Along with CO and likely other volatile molecules of weak dipole moment, this top ice layer is expected to host species with no dipole moment like N$_2$ \citep{Pontoppidan_2008}. This ice phase is thus weakly bound by intermolecular van der Waals forces and offers the possibility to desorb molecules by direct cosmic-ray impact \citep{Dartois2021,Huang_2020}, X-rays
\citep{Ciaravella_2012,Ciaravella_2016} in protoplanetary disks, or ultraviolet (UV) photons \citep{Oberg2007,Oberg2009,MunozCaro2010,Fayolle2011,Chen_2014,CruzDiaz2014}
generated by the interaction of cosmic rays with H$_2$ present in dense clouds \citep{Prasad1983, Cecchi-Pestellini_1992,Shen2004}. In the case of multicomponent ice mixtures, the UV-photodesorption yield of
CO is severely reduced in the presence of H$_2$O, also N$_2$
neighbours reduce the efficiency of CO UV-photodesorption
\citep{Bertin2012,Bertin_2013, Carrascosa2019}. But
photodesorption of CO molecules during UV-irradiation of pure CO ice is not much hindered by the presence of other species in the ice. This is due to the low efficiency in the formation of CO photoproducts during irradiation. 

The CO photodesorption yield reaches its highest value when this ice
is deposited at low temperatures (down to 7 K, the lowest temperature
studied experimentally) and decreases gradually at higher deposition
temperatures
\citep{Oberg2007,Oberg2009,MunozCaro2010,MunozCaro2016,Sie2022}. The
explanation for this phenomenon motivated further research. It was
found that the columnar structure of CO ice samples, grown at
incidence angles larger than 45$^{\circ}$, increases the effective ice
surface exposed to UV photons and therefore the photodesorption
efficiency \citep{GonzalezDiaz2019}, but ice surface effects cannot 
account for the large variations observed in the photodesorption 
of CO ice samples deposited at different temperatures
\citep{MunozCaro2016}. 
Absorption band shifts of CO ice in the UV and IR ranges only occurred
at deposition temperatures above 20 K \citep{Lasne2015,MunozCaro2016}, suggesting that CO ice grown at lower temperatures 
is amorphous below 20 K in our experiments, and therefore, the
decreasing photodesorption yield is not related to a transition from 
amorphous to crystalline ice, instead it might be associated to a
different degree of molecular disorder in CO ice samples, depending on 
their deposition temperature. Photon energy transfer via Wannier–Mott
excitons between the first photoexcited molecule in the ice and a
molecule on the ice surface capable to desorb was 
proposed \citep{Chen2017,MCCOUSTRA2018}.
Molecular disorder seems to enhance this energy transfer between
neighbor molecules. The color temperature variations measured at 
different deposition temperatures could also be the result of
molecular disorder \citep{Carrascosa_2021}. \cite{Urso2016},
\cite{Cazaux_2017}, and \cite{Carrascosa_2021} did not find
significant changes in the desorption behaviour or the color temperature 
of pure CO ice during controlled warm-up, which points to a low value 
of the diffusion in the ice. Finally, \cite{Sie2022} investigated the CO photodesorption yield dependence on ice thickness. 

CO ice density could also provide information about the ice structure 
dependence on deposition temperature. \cite{LUNA2022_aa} report the density of CO ice grown at different temperatures 
using laser interferometry and a microbalance. Here we use a different 
method to measure the density of CO ice grown at different
temperatures and confirm the results of 
\cite{LUNA2022_aa}; this method can be used by other authors to
estimate the ice density with the use of infrared spectroscopy, laser
interferometry and a quadrupole mass spectrometer. 

The most commonly used IR band strength
value of CO ice, 1.1 $\times$ 10$^{-17}$ cm molecule$^{-1}$, dates
from 1975 \citep{Jiang1975} and relies on much older references: They
adopted a density of 1.028 g cm$^{-3}$ for pure solid CO in the
${\alpha}$-phase (crystalline ice) at 30 K \citep{Vegard1930}, which
is significantly higher than the one reported by \cite{Roux1980}, 0.80
g cm$^{-3}$ at 20 K, and \cite{LUNA2022_aa}, 0.88 g cm$^{-3}$ at 20
K. \cite{Jiang1975} used a value of the refractive index, $n$ = 1.35,
that is also higher than those reported by \cite{Roux1980}, $n$ = 1.27
at 20 K, and \cite{LUNA2022_aa}, $n$ = 1.30 at 20 K. Finally,
\cite{Jiang1975} employed an ice thickness deposition rate between 0.5 and 2
$\mu$m min$^{-1}$, this rate is very high compared to the ones used in
modern setups devoted to astrochemistry and might affect the ice
structure; this issue is studied in the present article. For comparison, the highest deposition rate used in this
work was around 30 nm min$^{-1}$, which corresponds to a CO pressure 
of 10$^{-6}$ mbar during deposition.   
The IR band strength is a fundamental parameter needed to calculate the ice column density in the line of sight of the observations, it is therefore revisited in this paper.    

\section{Experimental} 
\subsection{Experimental setup} \label{sect:exp}

\begin{figure}
    \centering
    \includegraphics[width=\textwidth]{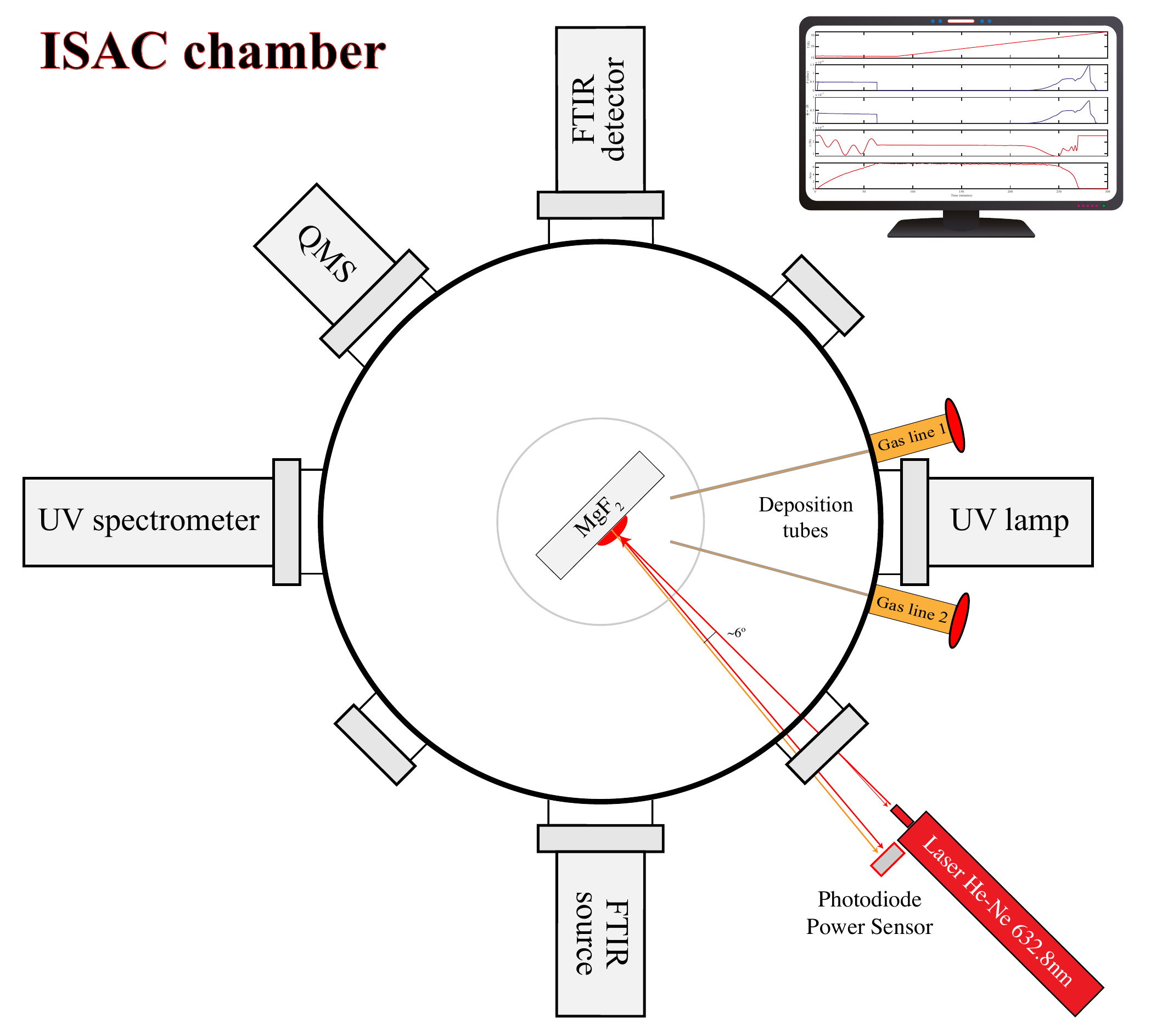}
    \caption{Intersection of ISAC at the level of the ice sample. Its multiple measurement components and sensors are shown, including the laser. QMS stands for quadrupole mass spectrometer, while FTIR is the Fourier transform infrared spectrometer with the FTIR source on one viewport and the FTIR detector on the opposite side. The UV spectrometer is located opposite to the vacuum-UV lamp.}
    \label{fig.ISAC_scheme}
\end{figure}

Experiments were carried out using the Interstellar Astrochemistry
Chamber (ISAC), described in more detail in \cite{MunozCaro2010}. ISAC
is an ultra-high vacuum (UHV) chamber with a base pressure of
4 $\times$ 10$^{-11}$ mbar designed to simulate the conditions present
in the interstellar medium (ISM), regarding temperature, pressure and
radiation field. A closed-cycle He cryostat allows to cool down the
tip of the cold finger to 8 K, where a sample holder with a MgF$_2$
window acting as the substrate for ice deposition is located. In the
experiments reported in this work the lowest achievable temperature
was 11 K because the radiation shield surrounding the sample holder
would block the laser signal and was removed for this reason. Temperature is
controlled by a Lakeshore temperature controller 331, with a precision
of 0.1 K. The gas line system in ISAC allows to introduce gas
species with a controlled composition, determined by quadrupole mass
spectrometry (QMS, Pfeiffer Vacuum, Prisma QMS
200). Fig. \ref{fig.ISAC_scheme} shows a scheme of the ISAC setup. The gas
line is connected to the main chamber through a leak valve. During
deposition, this valve is opened and the gas
is directed to the cold substrate via a deposition tube. The end of this
tube is about 3 cm  away from the substrate. The sample
holder can be rotated at any desired angle. Infrared spectra
were recorded during deposition, 
irradiation (if it is the case) and temperature programmed desorption
(TPD) using Fourier-transform infrared spectroscopy (FTIR) in
transmittance mode with a Bruker Vertex 70 at working spectral resolution of
2 cm$^{-1}$. Laser interferometry at 632.8 nm 
was implemented in ISAC to measure changes in the ice thickness for
this work. A He-Ne red laser (5.0 mW, 500:1 Polarization. Longitudinal mode frequency 
is $\sim$ 438 MHz with the spectral bandwidth being approximately 1400 MHz, Model: N-LHP-151) and a Silicon Photodiode Power Sensor to measure the
optical power of the laser light (model S120C) are placed with an
angle of $\sim6^{\circ}$. The distance of 40 cm between
the laser and the substrate allows to separate the laser reflection
coming from the viewport of ISAC and the laser reflection arriving
from the cold substrate where the ice sample is grown.

In a typical experiment, once the deposition temperature was reached, CO was introduced in the main chamber for 30 min by opening the leak valve of gas line 2 in 
Figure \ref{fig.ISAC_scheme} and the desired
deposition pressure of 1 $\times$ 10$^{-6}$ mbar was
reached. Additional experiments were performed using a lower CO
pressure in the 10$^{-8}$--10$^{-6}$ mbar range and longer deposition
times, to explore the effect
of deposition pressure in the ice density. The laser was turned on for
more than 30 min before starting the ice deposition, to ensure the
stability of the signal. Infrared spectra at 45$^{\circ}$ incidence of
the beam relative to the substrate were taken every 60 s during
deposition of the ice. The laser was pointing to the substrate and  
laser interference was measured. Purity of the ice
samples was >99.9 \%, 
as inferred from QMS data taken during deposition of the ice. Warm-up
of the samples after deposition was carried out at 0.1 K min$^{-1}$ until desorption ended near 30 K, and IR measurements were performed every 60 s.

\begin{figure}
    \centering
    \includegraphics[width=\textwidth]{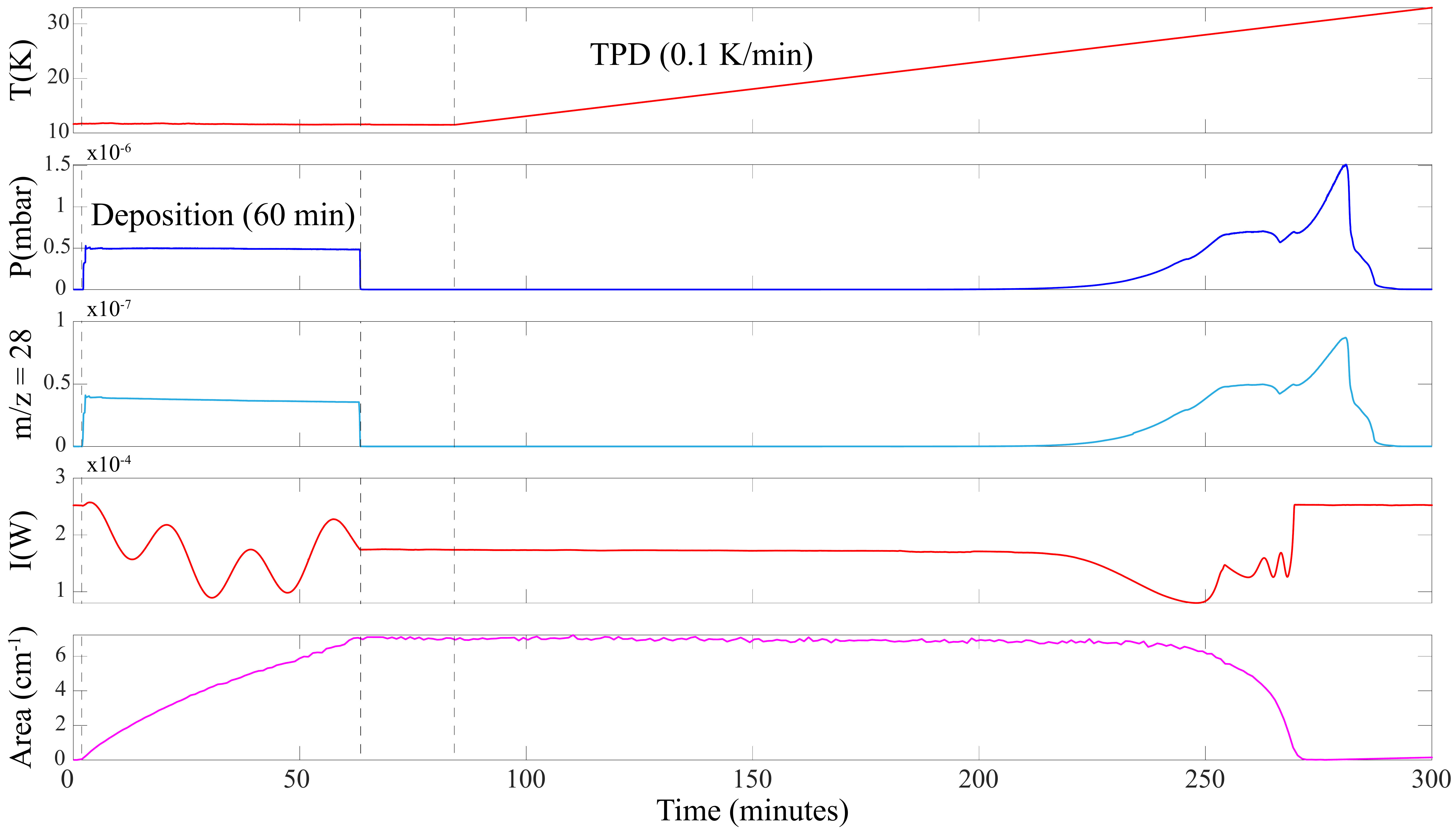}
    \caption{Main parameters controlled during experiments as a function of time. From top to bottom: temperature, pressure in the main chamber, $\frac{m}{z} = 28$ intensity from QMS data, laser intensity, integrated infrared absorbance from the CO ice band centered at 2138 cm$^{-1}$ that is proportional to its column density.}
    \label{fig.resumen}
\end{figure}

Fig. \ref{fig.resumen} shows the main parameters monitored during the experiments using various instruments. The temperature of the ice sample was measured using a silicon diode sensor placed just below the ice sample and attached to the sample holder. 
Pressure was monitored in the main chamber of ISAC using a Bayard-Alpert gauge located about 23 cm below the plane where deposition takes place. The relative sensitivity of Bayard-Alpert gauge to CO is 1.0, the same as N$_2$ \citep{Bayard20200}. Because CO was the only gas species deposited on the cold substrate, the pressure profile coincides with the $\frac{m}{z} = 28$ signal measured by QMS. Laser interference was measured continuously, creating an interference pattern during deposition of the ice, and a new interference during TPD. In this configuration, the deposited ice film is homogeneous according to ballistic simulations, in particular the formation of a columnar structure in the ice becomes only important at larger deposition angles, i.e. above 45$^{\circ}$ \citep{GonzalezDiaz2019}. Finally, the column density of the ice was obtained from IR spectroscopy using the areas that result from integration of the absorption bands, this is discussed in Sect. \ref{sect.band_strength}. 
The lower sensitivity of our FTIR and the time lapses between IR spectra may account for the delayed onset of thermal desorption of the ice compared to the other techniques presented in Fig. \ref{fig.resumen}. From this figure, the end of thermal desorption is observed simultaneously by IR spectroscopy and laser interferometry. The pressure gauge and the QMS still observe an increase of CO molecules in the gas due to desorption from CO ice layers accreted on the cold finger outside the MgF$_2$ substrate window. This effect is enhanced by the absence of a radiation shield in our laser interferometry experiments.

\subsection{Ice thickness estimation from laser interference} 
\label{sec:theo}
\begin{figure}
    \centering
    \includegraphics[width=1\textwidth]{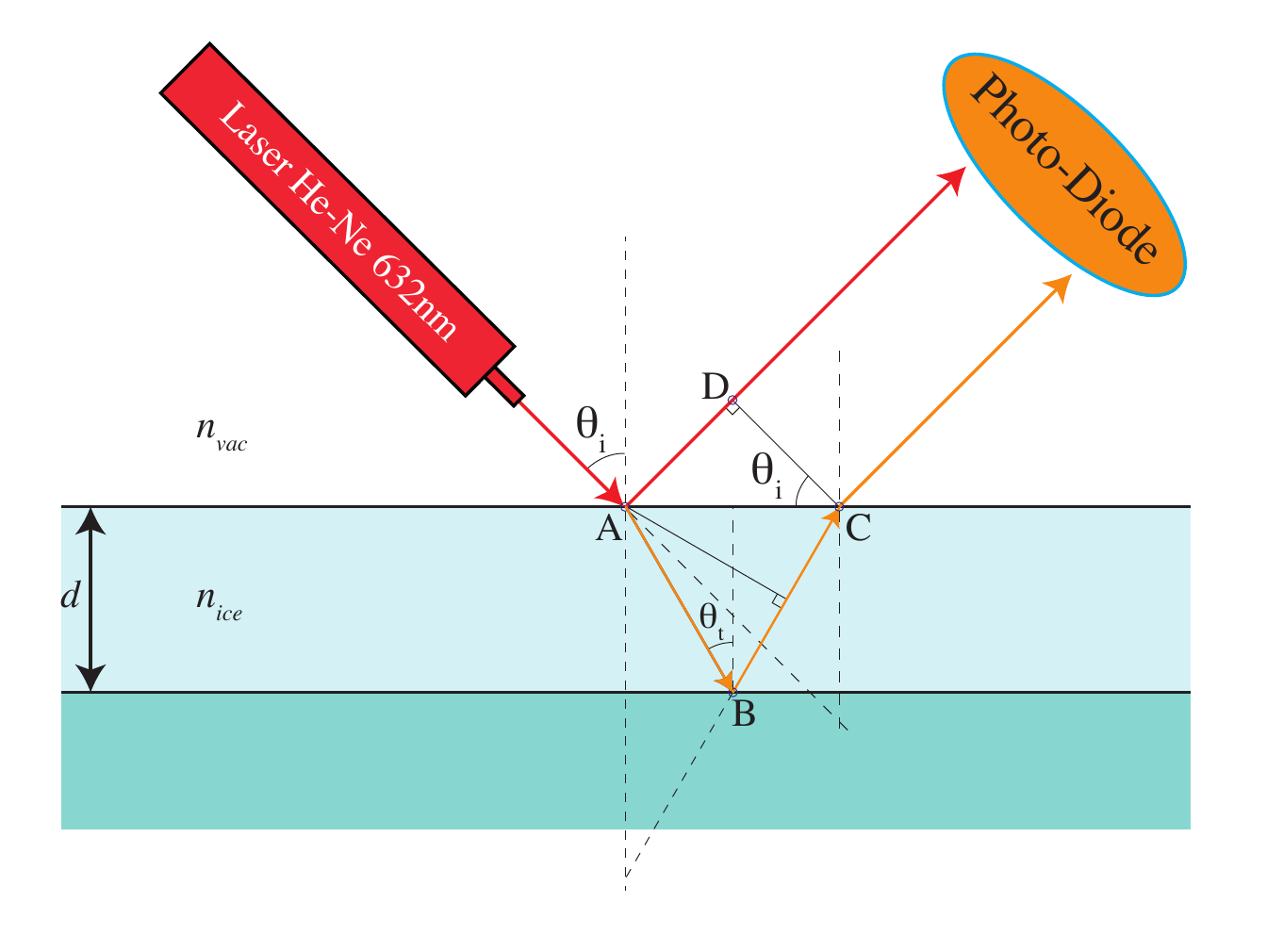}
    \caption{Sketch of the thin film interference principle, with the two reflections shown.}
    \label{fig:thinfilm}
\end{figure}

 The ice thickness was measured using single laser interference. One wave is reflected from the ice surface and one transmitted through the ice and then reflected from the surface of the cold substrate used for ice deposition. The reflection of both waves, shown in Figure \ref{fig:thinfilm}, creates an interference pattern as the result of ice thickness increase during deposition. If the deposition rate is constant, this interference forms a periodic pattern that allows estimation of the ice thickness, e.g. \cite{Hecht}.

The ray reflected on the ice surface (depicted in red in Fig. \ref{fig:thinfilm}) travels from $A$ to $D$, henceforth $|AD|$. It goes across a medium (vacuum in our experiments) with index of refraction $n_{vac}$, leading to optical path $n_{vac}|AD|$. The second ray travels across the ice and follows the path $n_{ice}(|AB|+|BC|)$ where $n_{ice}$ is the ice index of refraction. The optical path length difference of the reflected rays is
\begin{equation}
    \Lambda = n_{ice}(|AB|+|BC|) - n_{vac}|AD|
\end{equation}

From the triangle formed by $ABC$ it follows that
\begin{equation}
    |AB| = |BC| = \frac{d}{\cos{\theta_t}}
\end{equation}

The angle between $CD$ and the cold substrate is equal to $\theta_i$, see Fig. \ref{fig:thinfilm}, and therefore
\begin{equation}
    |AD| = |AC| \sin{\theta_i}
\end{equation}
$|AC|$ can be written as
\begin{equation}
    |AC| = 2d \tan{\theta_t}
\end{equation}

Using Snell's law,
\begin{equation}
    n_{vac} \sin{\theta_i} = n_{ice} \sin{\theta_t}
\end{equation}
we obtain 
\begin{align}
    |AD| &= 2d \tan{\theta_t} \frac{n_{ice}}{n_{vac}} \sin{\theta_t} \\ \indent
    & = 2d \frac{n_{ice}}{n_{vac}} \frac{\sin^2{\theta_t}}{\cos{\theta_t}}
\end{align}

The optical path length difference can now be expressed as

\begin{align}
    \Lambda &= \frac{2 n_{ice} d}{\cos{\theta_t}} - 2d n_{vac} \frac{n_{ice}}{n_{vac}} \frac{\sin^2{\theta_t}}{\cos{\theta_t}} \\ \indent
    &= \frac{2 n_{ice} d}{\cos{\theta_t}}(1 - \sin^2{\theta_t}) \\ \indent
    &= 2 n_{ice} d \cos{\theta_t}
\end{align}

The phase difference between the two reflected beams is $k_0 \Lambda$ where $k_0=\frac{2\pi}{\lambda_0}$ is the free-space propagation number. Adding the relative phase shift of $\pm \pi$ radians between the reflected beams we obtain
\begin{align}
    \delta &= k_0 \Lambda \pm \pi \\ \indent
    &= \frac{4 \pi n_{ice}}{\lambda_0} d \cos{\theta_t} \pm \pi \label{eq:phaseshift}
\end{align}

Using Snells´s law, eq. \ref{eq:phaseshift} can also be expressed as
\begin{equation}
    \delta = \frac{4\pi d}{\lambda_0}\sqrt{n_{ice}^2 - n_{vac}^2 \sin^2{\theta_i}} \pm \pi
\end{equation}

The sign of the phase shift has no physical meaning, the minus sign is selected for convenience. An interference maximum occurs when the phase shift is $\delta$ = $2m\pi$ where $m$ = 0, 1, 2, ..., then 
\begin{equation}
    d \cos{\theta_t} = (2m + 1)\frac{\lambda_{ice}}{4} \label{eq:max}
\end{equation}
where $\lambda_{ice} = \frac{\lambda_0}{n_{ice}}$. A minimum in the interference curve corresponds to  reflected curves displaying peaks of opposite signs, i.e. $\delta$ = $(2m + 1)\pi$, then
\begin{equation}
    d \cos{\theta_t} = 2m \frac{\lambda_{ice}}{4} \label{eq:min}
\end{equation}

In practice, we estimated the ice thickness deposited during the time lapse between two consecutive minima as
\begin{equation}
    d = \frac{\lambda_{0}}{2 n_{ice} \cos{\theta_t}} \label{eq:min_practice}
\end{equation}

where $n_{ice}$ depends on the ice deposition temperature, according to \cite{LUNA2022_aa}, $\lambda_{0}$ = 632.8 nm for the He-Ne laser, and the angle ${\theta_t}$ is 

\begin{equation}
\theta_t = \arcsin{(\frac{ {n_{vac}} \sin{\theta_i}}{n_{ice}})} \label{eq:angle}
\end{equation}

with ${\theta_i}$ = 3$^{\rm o}$. 

Finally, the total reflected signal collected by the sensor in our experiments is the result of the interference of the two reflected rays. If for simplicity only the first back-reflection is considered, what we obtain is known as the reflexibility $R$ \citep{Ishikawa}. If the medium surrounding the ice layer is air or vacuum, $n_{vac}$ = 1, then

\begin{equation}
    R = 2r^2 \frac{1 - \cos{(2\delta)}}{1 + r^2 (r^2 - 2\cos{(2\delta)})}
    \label{eq:R}
\end{equation}

where 
\begin{equation}
    \delta = \frac{2 \pi n_{ice}}{\lambda_0} d \cos{\theta_t} \label{eq:phaseshift2}
\end{equation}

\begin{equation}
    r = \frac{\sin{(\theta_t - \theta_i)}}{\sin{(\theta_t + \theta_i)}}
\end{equation}

and 
\begin{equation}
    n_{ice} = \frac{\sin{(\theta_i)}}{\sin{(\theta_t)}}
\end{equation}

Equation \ref{eq:R} was used to fit the interference curves of the measured samples as can be seen in Fig. \ref{fig:laser_exp}.

\subsection{Infrared band strength of CO ice} \label{sect.band_strength}

The column density $N$ of the ice layer accreted on the cold substrate, in molecule cm$^{-2}$, is obtained from integration of the infrared absorption band\\
\begin{equation}
 N = \frac{1}{\mathcal{A}}\int_{band}\tau_{\nu}d{\nu}
\label{eq:N}
\end{equation}
where ${\tau}_{\nu}$ is the optical depth of the band, $d\nu$ the wavenumber differential in cm$^{-1}$, and $\mathcal{A}$ the band strength in cm molecule$^{-1}$. The integrated absorbance, $A_{int}$ is 
equal to 0.43 $\times$ $\tau$, where $\tau$ is the integrated optical 
depth of the band. The most commonly adopted band strength of CO ice is $\mathcal{A}$(CO)=1.1 $\times$ 10$^{-17}$ cm molecule$^{-1}$ \citep{Jiang1975}. This value of $\mathcal{A}$(CO) is revisited in Sect. \ref{sect:band_strength_estimation}. 

\section{Results} \label{sect:results}

\subsection{Interference of substrate window used for ice deposition} 

\begin{figure}
    \centering
    \includegraphics[width=\textwidth]{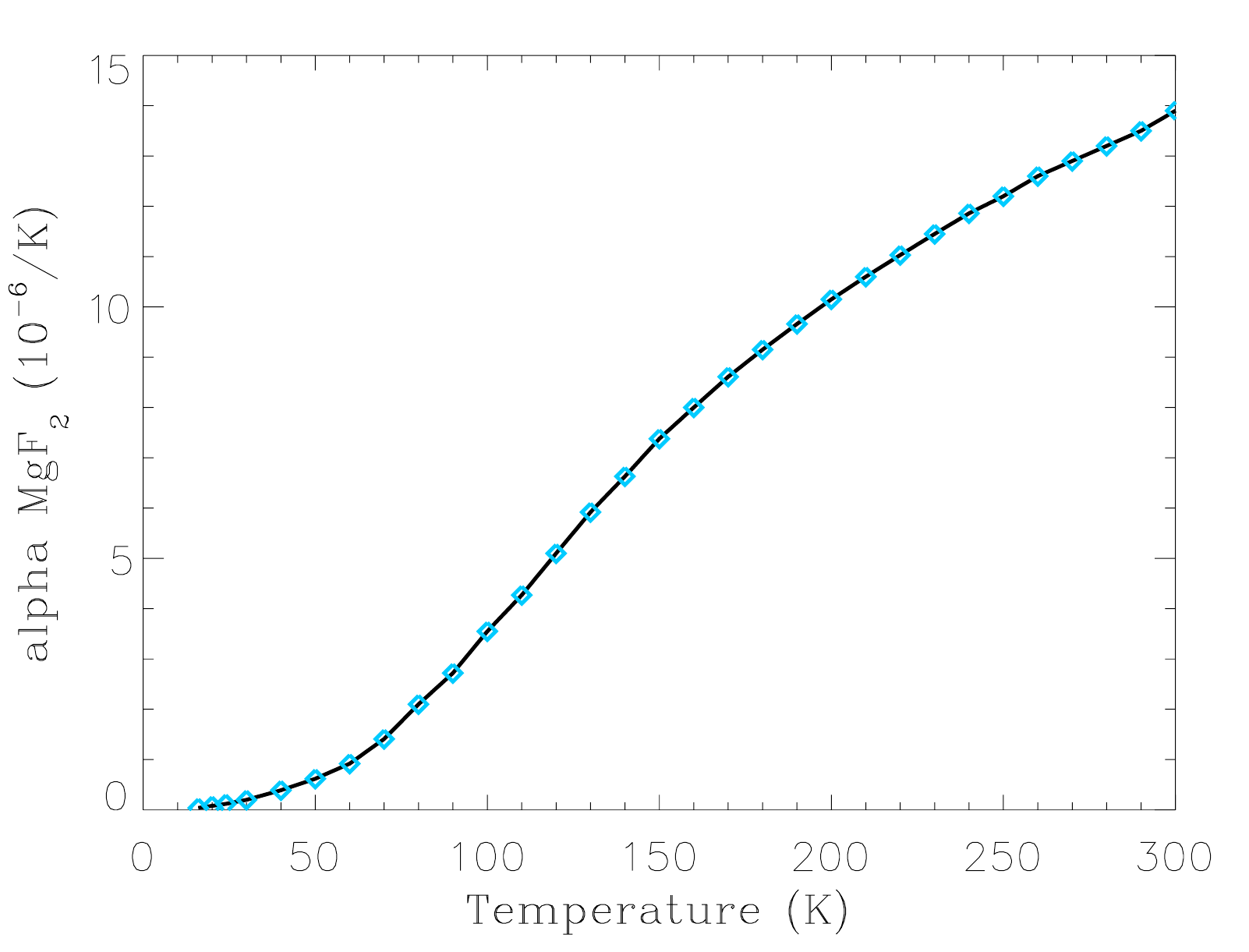}
    \caption{Expansion coefficient $\alpha$ of MgF$_2$ obtained from data reported in the literature. Data points are represented as blue empty squares. The solid black line was obtained from interpolation of these data points.}
    \label{fig.MgF2_expansion_alpha}
\end{figure}

\begin{figure}
    \centering
    \includegraphics[width=\textwidth]{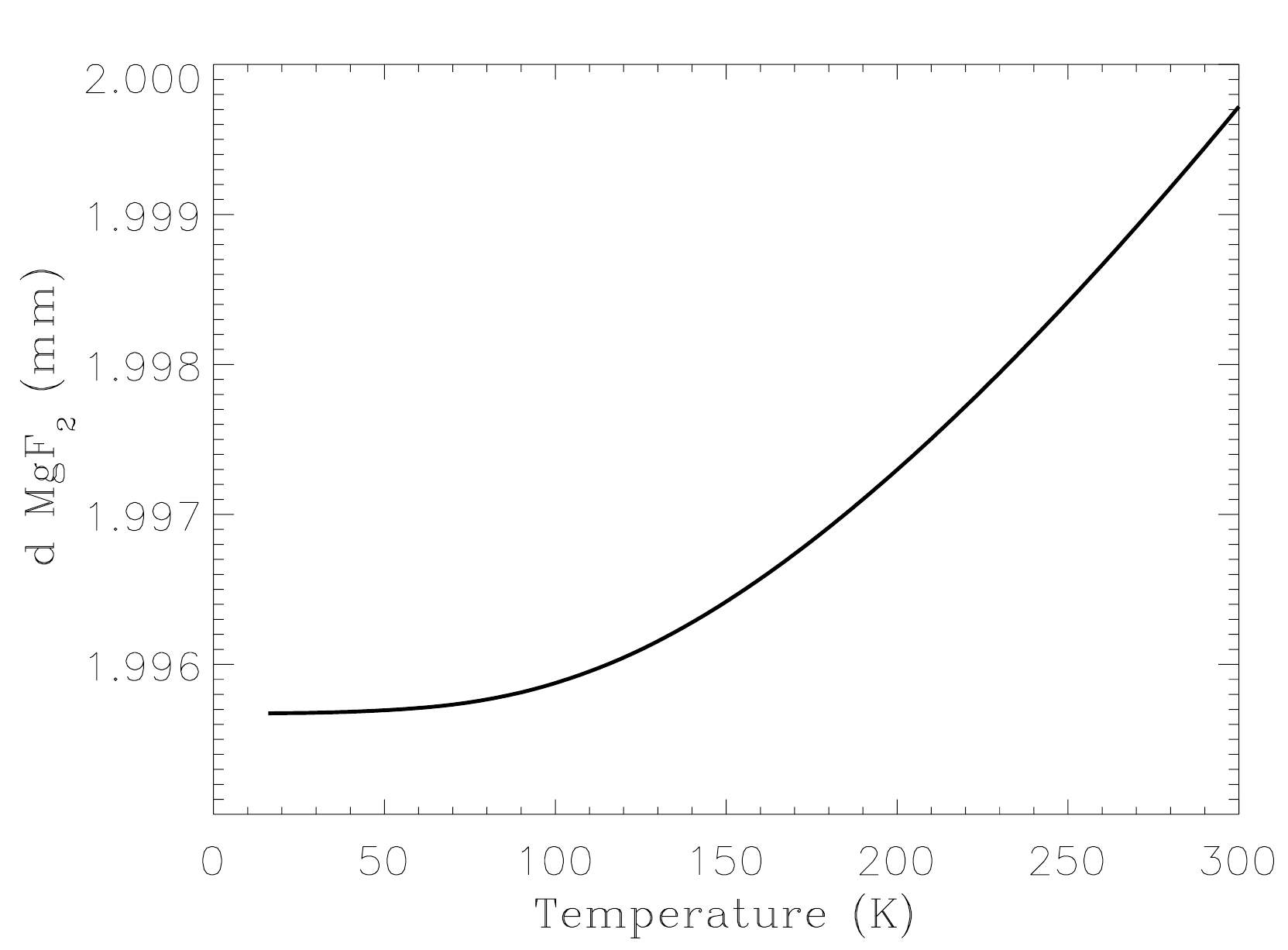}
    \caption{Estimated thickness $d$ (in mm) of the 2 mm-thick MgF$_2$ window, used as the substrate for CO ice deposition, during TPD experiments.}
    \label{fig.MgF_expansion_d}
\end{figure}

The MgF$_2$ window used for ice deposition contracts and expands during cool down and warm-up, respectively. This can lead to a laser interference pattern that may be difficult to disentangle from the one caused by the ice layer during warm-up. The MgF$_2$ window thickness does not change appreciably during ice deposition because the temperature was kept constant. What follows is an estimation of the thickness variation of the 2 mm-thick MgF$_2$ window used in the reported experiments during warm-up or cool down. Figure \ref{fig.MgF2_expansion_alpha} represents the expansion coefficient $\alpha$ of MgF$_2$ at different temperatures based on literature data values \citep{BROWDER1975,Browder1977,Feldman1975O} for the 632.8 nm wavelength (blue empty squares). The solid black line is obtained from interpolation of these data points. Using the interpolated values and formula
\begin{equation}
d(t_2) = d(t_1) \cdot ( 1 + \alpha(t_2) \cdot (T(t_2)-T(t_1)) )  
\end{equation}

where $t_1$ and $t_2$ are the time values for substrate temperatures $T(t_1)$ and $T(t_2)$, the MgF$_2$ window thickness d is calculated at the different temperatures as shown in Fig. \ref{fig.MgF_expansion_d}. This figure shows that during the TPD of CO ice, which desorbed around 30 K in our experiments, there is no expansion of the MgF$_2$ window and ice thickness variations are therefore responsible for the interference curve measured during TPD experiments. For more refractory ices the expansion of the MgF$_2$ window during the TPD needs to be taken into account. In particular, the TPD of water ice from 10 K to its desorption near 170 K, leads to an expansion of the MgF$_2$ window of about one micron. 

A similar calculation was performed for a KBr window substrate. For this material, the literature data \citep{Feldman1975O,Meincke1965} only allowed to calculate the thickness expansion above 90 K, an increase of 50 K in temperature corresponds to about 4 $\mu$m expansion of the 2 mm-thick KBr substrate. 

For this reason, we selected the MgF$_2$ window as the substrate for laser interferometry measurements of ice samples.

\subsection{Density measurements}
The volumetric density, $\rho$, of an ice layer of mass $M$, thickness $d$ and flat surface area $A$ = 1 cm$^2$ can be expressed as
\begin{equation} \label{eq:dens}
    \rho = \frac{M}{d \cdot A} = \frac{ m_{CO} \cdot n_{CO}}{N_A \cdot d \cdot A} = \frac{m_{CO} \cdot N}{N_A \cdot d}
\end{equation}
where $m_{CO}$ = 28 g is the molar mass of CO, Avogadro's number $N_A$ = 6.022 $\times$ 10$^{23}$ molecules, and the ice column density measured in the infrared $N$ is the number of molecules $n_{CO}$ per cm$^2$. Estimation of the density using this method involves measurement of $d$ using laser interferometry and the value of $N$ from infrared spectroscopy in transmittance using Eq. \ref{eq:N}. The latter commonly assumes a value of the band strength   $\mathcal{A}(CO)$ = 1.1 $\times$ 10$^{-17}$ cm molecule$^{-1}$ \citep{Jiang1975}. 
We developed a second method to obtain the density ($\rho$) of CO ice samples and provide a new estimation of $\mathcal{A}(CO)$. 
The general expression for the total number of molecules that impinge on the cold substrate per unit time (s) and unit area (cm$^2$) is known as the collision frequency per unit area ($Z_w$), this is 
\begin{equation} \label{eq:flow1}
Z_w = \frac{1}{4} n \overline{v} 
\end{equation}

where $n$ is the average number of molecules per unit volume and $\overline{v}$ is the average velocity of the molecules according to classical statistical mechanics. For a diluted gas that follows the Maxwell distribution this becomes

\begin{equation} \label{eq:flow2}
Z_w = \frac{P}{ \sqrt{2 \pi m k T_g} } 
\end{equation}

 where $P$ is the gas pressure in the chamber, $m$ is the molecular mass, $k$ is the Boltzmann constant and $T_g$ is the gas temperature. The validity of the Maxwell distribution in surface physics experiments has been tested (\cite{Todorov2017}, and ref. therein). In the ultra-high vacuum conditions of our chamber the background pressure $P_0$ is in the 10$^{-11}$ mbar range. Due to the flow of CO molecules entering the ISAC chamber, the pressure increase measured during deposition in the reported experiments, $\Delta P$, ranged from 1 $\times$ 10$^{-6}$ to 1 $\times$ 10$^{-8}$ mbar, and therefore $P$ $\approx$ $\Delta P$. An analog formula was used by \cite{SCHUTTE1993_Icarus} to estimate the flow in ice deposition experiments, this is
 
\begin{equation} \label{eq:flow3}
Z_w = C^{\prime} \frac{\Delta P}{ \sqrt{m T_g} } 
\end{equation}

where $C^{\prime}$ depends on the efficiency of the vacuum pump and the geometry of the system.

\begin{figure}
    \centering
    \includegraphics[width=\textwidth]{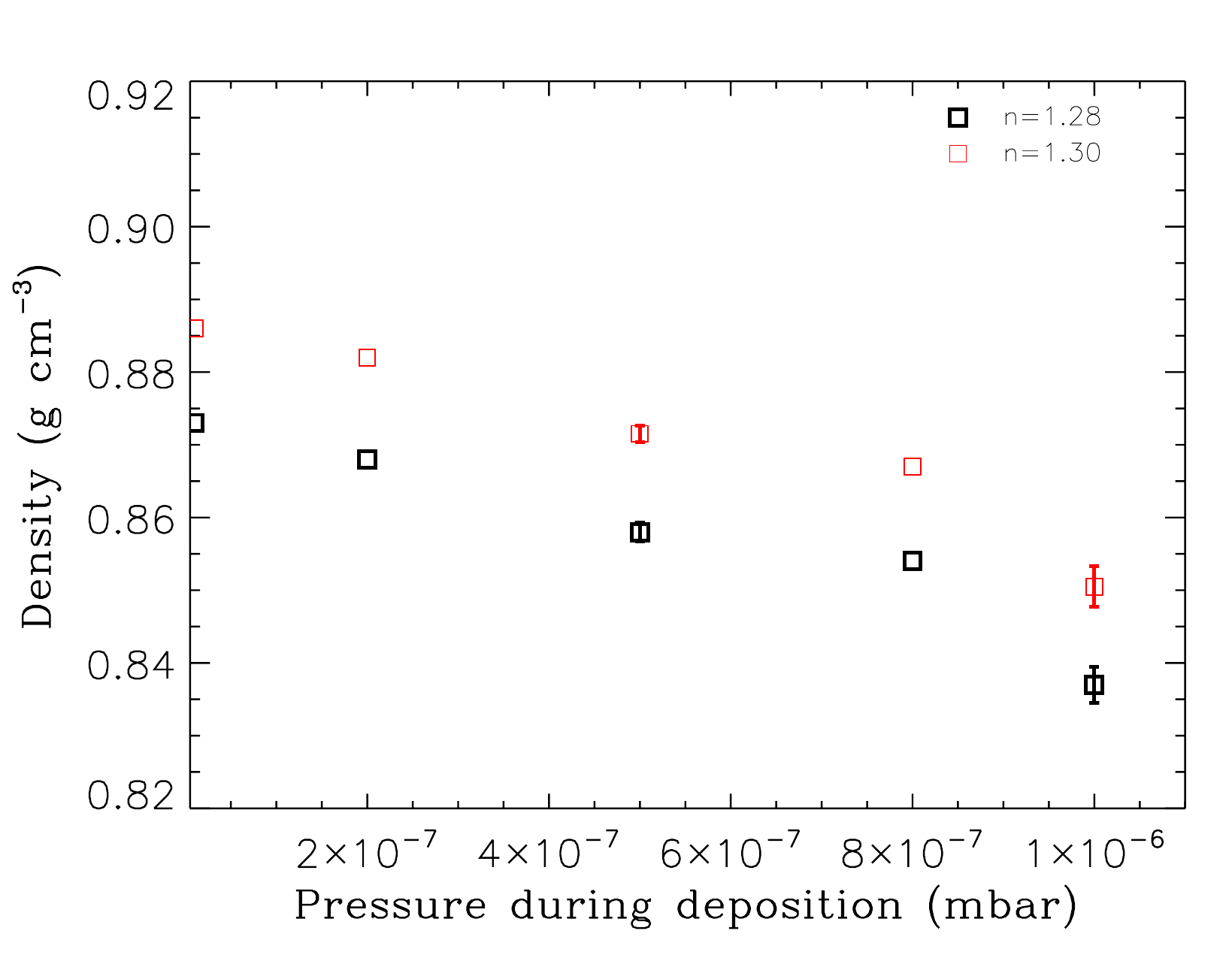}
    \caption[]{Density of CO ice deposited at different CO gas pressure. Black squares correspond to the density estimated using a refractive index value of $n$ = 1.28 from \cite{LUNA2022_aa} for 13 K, a temperature close to our deposition temperature of 11 K in these experiments. Red squares correspond to density values using $n$ = 1.30 as the typical value of crystalline CO ice grown above 20 K in \cite{LUNA2022_aa}.}
    \label{fig.CO_density_vs_deltaP}
\end{figure}

Therefore, the total number of molecules that {\it accrete} on the cold substrate per cm$^2$ corresponds to the ice column density $N$ and is obtained from 

\begin{equation} \label{eq:N_f}
N = C \cdot \Delta P \cdot t 
\end{equation}
where the proportionality factor $C$ must include the sticking probability $S$. This value of $C$ is $C = \frac{S}{\sqrt{2 \pi m k T_g}}$ for the Maxwell distribution or 
$C = S \frac{C^{\prime}}{ \sqrt{m T_g} } $ according to \cite{SCHUTTE1993_Icarus}.
Eq. \ref{eq:N_f} simply shows that the column density of the accreted ice layer at deposition time $t$ is proportional to the increase in pressure $\Delta P$. This condition was fulfilled in our experiments: for the same deposition temperature, the deposition rate expressed in units of ice thickness (nm) per second increased linearly with $\Delta P$; it is shown below that very small deviations were found attributable to a small dependence of ice density with $\Delta P$. The CO gas temperature $T_g$ is not expected to vary appreciably between experiments. The sticking probability $S$ is close to unity and, within experimental errors, does not vary as a function of substrate temperature during deposition until this temperature approaches the desorption temperature (e.g., \cite{Sandford_Allamandola_1988,Gerakines1995A&A,Bisschop2006}). We thus considered $C$ as a constant in our experiments. 

\begin{figure}
    \centering
    \includegraphics[width=\textwidth]{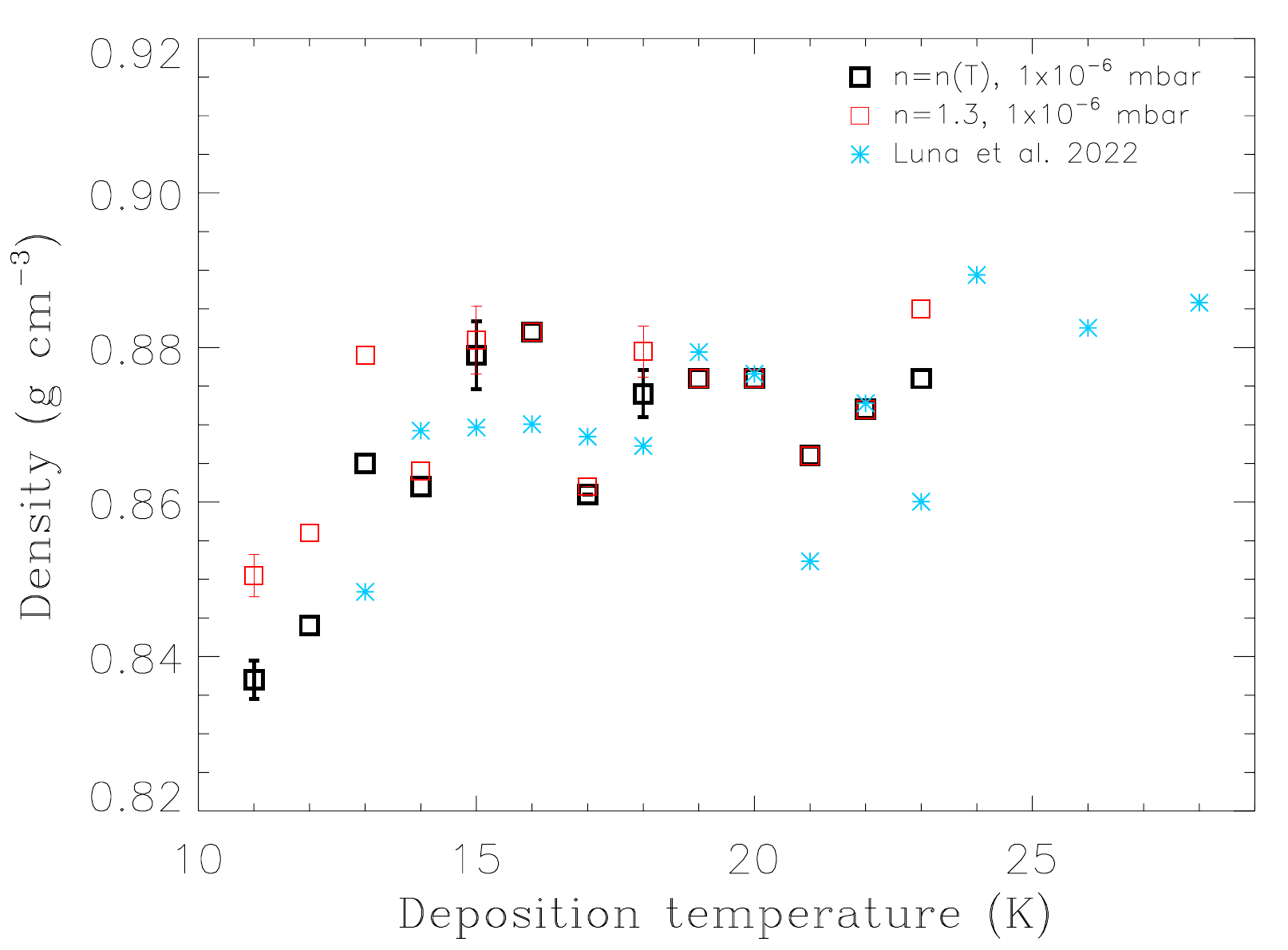}
    \caption[]{Density of CO ice deposited at different temperatures. Black squares correspond to densities estimated using refractive index values at the different deposition temperatures provided in \cite{LUNA2022_aa}, $n$=$n$($T$). Red squares correspond to densities estimated using a constant value of the refractive index, $n$ = 1.3 for all deposition temperatures. For comparison, the density values reported in \cite{LUNA2022_aa} are represented as blue stars. In all the experiments, the CO pressure during deposition was 1 $\times$ 10$^{-6}$ mbar.}
    \label{fig.CO_density_vs_T}
\end{figure}

The density value of 0.876 g cm$^{-3}$ for CO ice deposited at 20 K, and the refractive indices of CO ice samples deposited at 13 to 28 K  \citep{LUNA2022_aa} were adopted. From eqs. \ref{eq:dens} and \ref{eq:N_f}, the density of ice deposited at temperature $T$ different from 20 K can be obtained using
\begin{equation} \label{eq:density}
\rho(T) = \frac{\Delta P(T) \cdot t(T)}{d(T)} \cdot \rho(20 K) \frac{d(20 K)}{\Delta P(20 K) \cdot t(20 K)}
\end{equation}
We note that the factor $C$ in eq. \ref{eq:N_f} does not appear in eq. \ref{eq:density} and therefore our estimation of the ice density at the different deposition temperatures is independent from the value or the expression adopted for $C$. This is a great advantage because $C$ is difficult to measure in practice. This method thus requires the measurement of $\Delta P(T)$ and the ice thickness $d(T)$ during deposition in the reference experiment of known density, here the 20 K deposition temperature, and a second experiment at different deposition temperature with unknown density.
The parameters of the experiments are summarized in Table \ref{table:log}. 
The first and second column are the substrate temperature, $T_{dep}$, and $\Delta P$ during deposition. The third column is the term $\frac{\Delta P(T) \cdot t(T)}{d(T)}$, in mbar s m$^{-1}$, in eq. \ref{eq:density}. The fourth column is the ice density $\rho$ obtained with this formula, where the refractive index $n$ values, used for the estimation of the ice thickness $d$, are those reported in \cite{LUNA2022_aa} and vary with deposition temperature. To test the effect of these $n$ values in the density estimation, the fifth and sixth columns are $\frac{\Delta P(T) \cdot t(T)}{d(T)}$ and $\rho$ for the fixed value of n = 1.30 at all temperatures, this is the $n$ value for the 20 K deposition experiment. The last column is the total area $(\Delta P(T) \cdot t(T))_{TPD}$ of the TPD during desorption divided by the same parameter for the deposition, $(\Delta P(T) \cdot t(T))_{dep}$. For experiments performed at the same deposition temperature, this value was highly reproducible and therefore a significant decrease at a certain deposition temperature was indicative of a decrease in the effective sticking probability S during deposition: this effect became observable at 24 K, in agreement with \cite{Sandford_Allamandola_1988}, and was important at 26 K, where this decrease was near 20 \%. For this reason, our method does not allow to estimate the ice density at deposition temperatures above 23 K unless the value of S is known at these high temperatures.

Figure \ref{fig.CO_density_vs_deltaP} presents the estimated density of CO ice deposited at 11 K in various experiments performed at different deposition pressure of CO gas. Error bars correspond to the standard deviation calculated from at least two repeated experiments performed at the same conditions. According to this data, a relatively small increase in the CO ice density is observed when the deposition was done at the lower CO pressures. 

Figure \ref{fig.CO_density_vs_T} shows the CO ice volumetric densities estimated for the various deposition temperatures from 11 to 23 K. As already mentioned in 
Sect. \ref{sect:exp}, the reported experiments were performed with no radiation shield and therefore 11 K, instead of the usual 8 K in our ISAC setup, was the lowest achievable temperature. Error bars correspond to the standard deviation calculated from at least two repeated experiments performed at the same conditions. A good agreement is found with the recent values reported in \cite{LUNA2022_aa}, which is reproduced here for comparison (blue stars). Above 14 K, the CO ice density is stabilized around 0.88 g cm$^{-3}$. A decrease in the density can be appreciated below 13 K deposition temperature.

\begin{table}
\begin{normalsize}
\caption{Experimental parameters \label{table:log}}
\begin{tabularx}{\textwidth}{*{7}{X}} 
$T_{dep}$ & $\Delta P$ & $\frac{\Delta P(T) \cdot t(T)}{d(T)}$ & $\rho(T)$ & $\frac{\Delta P(T) \cdot t(T)}{d(T)}$ $n=1.3$ & $\rho(T)$ ~~~~~~ $n=1.3$ & $\frac{(\Delta P(T) \cdot t(T))_{TPD}}{(\Delta P(T) \cdot t(T))_{dep}}$ \\
K & mbar & mbar s m$^{-1}$ & g cm$^{-3}$ & mbar s m$^{-1}$ &  g cm$^{-3}$ &  \\ 
\hline 
\hline
11 & 1 $\times$ 10$^{-6}$ & 1465.30 & 0.831 & 1488.23 & 0.844 & 1.16 \\
11 & 1 $\times$ 10$^{-6}$ & 1487.03 & 0.843 & 1510.31 & 0.857 & 1.15 \\ 
11 & 2 $\times$ 10$^{-7}$ & 1530.94 & 0.868 & 1554.90 & 0.882 & 1.21 \\ 
11 & 4 $\times$ 10$^{-7}$ & 1640.57 & 0.931 & 1666.24 & 0.945 & 1.19 \\
11 & 5 $\times$ 10$^{-7}$ & 1507.72 & 0.855 & 1531.32 & 0.869 & 1.19 \\
11 & 5 $\times$ 10$^{-7}$ & 1517.92 & 0.861 & 1541.68 & 0.874 & 1.17 \\
11 & 8 $\times$ 10$^{-7}$ & 1505.57 & 0.854 & 1529.12 & 0.867 & 1.19 \\
11 & 1 $\times$ 10$^{-8}$ & 1538.55 & 0.873 & 1562.63 & 0.886 & 1.30 \\ 
12 & 1 $\times$ 10$^{-6}$ & 1487.15 & 0.844 & 1510.42 & 0.856 & ~----- \\ 
13 & 1 $\times$ 10$^{-6}$ & 1525.01 & 0.865 & 1548.88 & 0.879 & 1.17 \\
14 & 1 $\times$ 10$^{-6}$ & 1519.76 & 0.862 & 1523.28 & 0.864 & 1.17 \\
15 & 1 $\times$ 10$^{-6}$ & 1568.16 & 0.889 & 1570.58 & 0.891 & 1.16 \\
15 & 1 $\times$ 10$^{-6}$ & 1532.77 & 0.869 & 1535.14 & 0.871 & 1.17 \\
16 & 1 $\times$ 10$^{-6}$ & 1554.15 & 0.882 & 1555.35 & 0.882 & 1.16 \\
17 & 1 $\times$ 10$^{-6}$ & 1517.99 & 0.861 & 1520.33 & 0.862 & 1.18 \\
17 & 1 $\times$ 10$^{-6}$ & 1518.01 & 0.861 & 1520.36 & 0.862 & 1.16 \\ 
18 & 1 $\times$ 10$^{-6}$ & 1553.39 & 0.881 & 1563.02 & 0.887 & 1.16 \\ 
18 & 1 $\times$ 10$^{-6}$ & 1528.33 & 0.867 & 1537.81 & 0.872 & 1.17 \\
19 & 1 $\times$ 10$^{-6}$ & 1544.58 & 0.876 & 1544.58 & 0.876 & 1.15 \\
20 & 1 $\times$ 10$^{-6}$ & 1544.41 & 0.876 & 1544.41 & 0.876 & 1.16 \\
20 & 1 $\times$ 10$^{-6}$ & 1542.38 & 0.875 & 1542.38 & 0.875 & 1.16 \\ 
21 & 1 $\times$ 10$^{-6}$ & 1526.56 & 0.866 & 1526.56 & 0.866 & 1.17 \\
22 & 1 $\times$ 10$^{-6}$ & 1537.54 & 0.872 & 1537.54 & 0.872 & 1.16 \\
23 & 1 $\times$ 10$^{-6}$ & 1543.98 & 0.876 & 1560.21 & 0.885 & 1.14 \\
\hline
\end{tabularx}
\end{normalsize}
\end{table}

\subsection{Estimation of infrared band strength of CO ice} \label{sect:band_strength_estimation}

\begin{figure}
    \centering
    \includegraphics[width=1\textwidth]{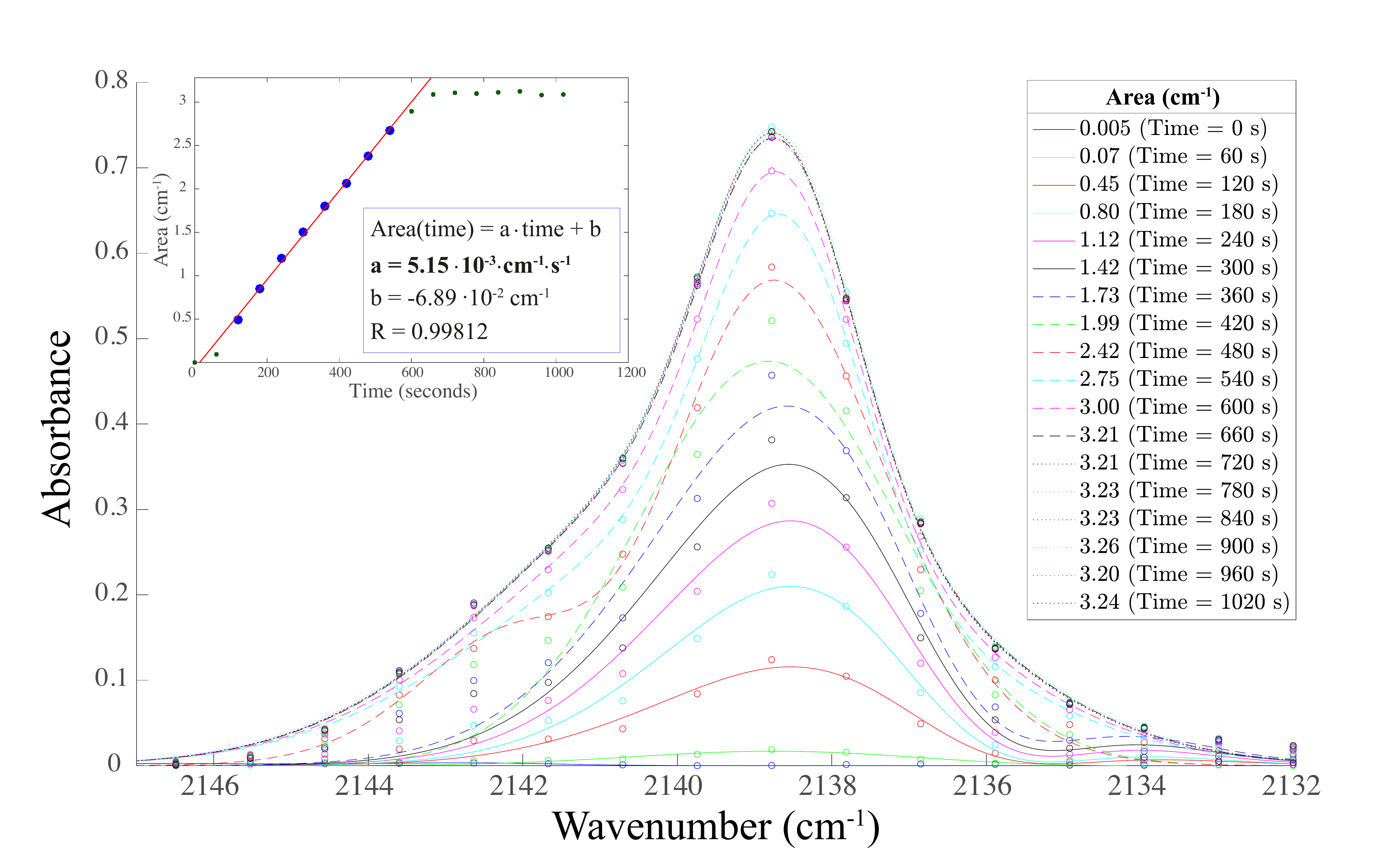}
    \caption{Infrared spectra of CO ice deposited at 20 K as a function of deposition time. Left inlet: Red trace is the fit of the integrated absorbance (¨Area (cm$^{-1}$)¨ in this figure) as a function of deposition time. 
    The goodness of fit is $R$ > 0.99. Right inlet: The values of the integrated absorption at various deposition times are provided.}
    \label{fig:A_int_exp}
\end{figure}

\begin{figure}
    \centering
    \includegraphics[width=1\textwidth]{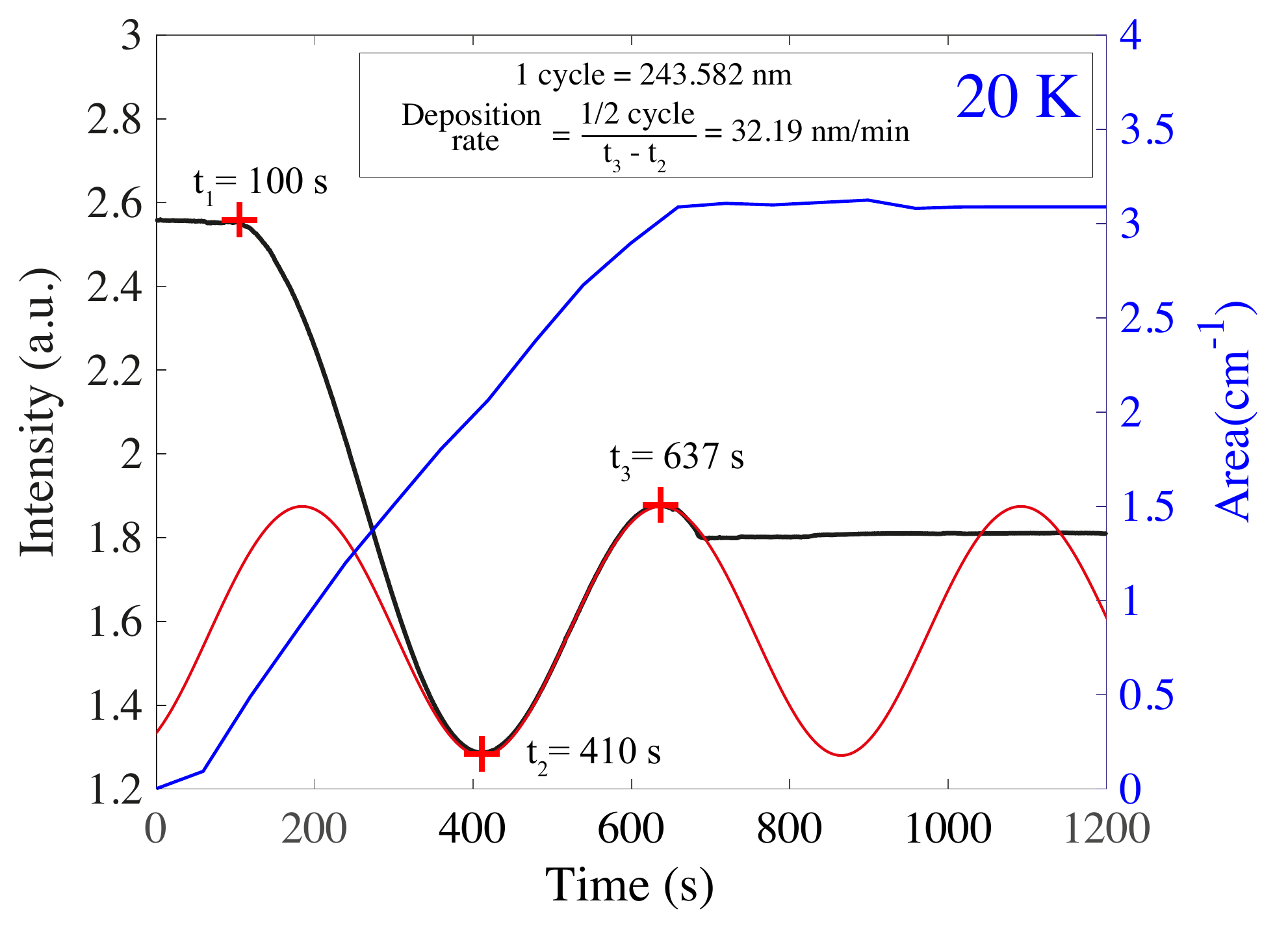}
    \caption{Relatively short ice deposition at 20 K to estimate the infrared band strength using $d$ value from laser interference shown in figure (black trace). Red trace is the fit of the second half of the cycle from 410 to 637 s to obtain the ice deposition rate in nm/min. Red crosses indicate the position of the minimum and maximum points of the cycle. \textbf{Blue trace corresponds to the integrated band area as a function of time during deposition of the ice, presented in Fig. \ref{fig:A_int_exp}.}}
    \label{fig:laser_exp}
\end{figure}

The infrared band strength of CO ice, $\mathcal{A}(CO)$, can be estimated from eqs. \ref{eq:N} and \ref{eq:dens} according to    
\begin{equation} \label{eq:band_stregth}
    \mathcal{A}(CO) = \frac{2.3 \cdot A_{int} \cdot m_{CO}}{N_A \cdot \rho \cdot d} 
\end{equation}
Eq. \ref{eq:band_stregth} was used for the estimation of the infrared band strength of CO ice, $\mathcal{A}(CO)$, at 20 K deposition temperature. The recent value of the density ($\rho$) at 20 K from \cite{LUNA2022_aa}, and the parameters measured in our experiments, i.e. ice thickness $d$ and integrated band strength $A_{int}$, were used as input in this equation. The main difficulty was the precise determination of the integrated band area, $A_{int}$, at different deposition temperatures. Indeed, laser interferometry allows to measure micron-thick ices, but due to saturation of the infrared absorption for thick ice depositions, only a few infrared spectra could be acquired within the duration of one laser interference cycle, i.e. the distance between two minima. Fig. \ref{fig:A_int_exp} shows a typical experiment, performed at 20 K, for the calculation of the integrated band areas as a function of deposition time. The laser interference corresponding to this experiment is shown in Fig. \ref{fig:laser_exp}, along with the integrated band areas from Fig. \ref{fig:A_int_exp}, deposition took place between $\sim$ 100 s and 637 s. Despite this limitation, no significant variations were found in the value of $\mathcal{A}(CO)$ as a function of deposition temperature. As mentioned in Sect. \ref{sect:exp}, the spectra were acquired at a 45$^{\circ}$ angle of the infrared beam with respect to the cold MgF$_2$ window where the ice was deposited. We repeated this experiment using the same procedure but this time the infrared spectra of the ice sample were taken at normal incidence angle between the infrared beam and the MgF$_2$ window and, from the ratio of the integrated absorbances, it was found that a multiplication factor of 1/1.30 $\approx$ 0.77 is required for the estimation of the column density measured at 45$^{\circ}$. This correction accounts for the larger pathlength of the infrared beam across the ice layer in the 45$^{\circ}$ configuration. The band strength value obtained for 3 experiments where CO ice samples were grown at 20 K is

\begin{equation} 
\mathcal{A}(CO) = (8.7 \pm 0.5)  \times 10^{-18} ~ {\rm cm ~ molecule}^{-1}. 
\end{equation}
This value is applicable for infrared spectra measured at normal
incidence angle. 

\section{Conclusions} \label{Conclusions}

We found a good agreement between the volumetric ice densities
obtained by \cite{LUNA2022_aa} at different deposition temperatures of CO
and those reported in this work. The different density values reported here for varying CO pressure during deposition indicate that this parameter must also be considered when data among different laboratories are compared.
This new methodology can thus be used for calculation of the ice
density at different ice accretion temperatures before the onset of
thermal desorption, provided that this value is known for one specific
temperature. This method presents some advantages: i) only one laser,
instead of two, is required, to measure the ice thickness, ii) the
microbalance is replaced by the more commonly used FTIR spectrometer
for estimation of the ice mass, this allows estimation of the IR band
strength of the ice and monitoring of possible structural ice changes, 
and iii) the pressure of the vacuum chamber is usually recorded during the experiments, or alternatively the mass
spectrum of the deposited species, both techniques are routinely used in experimental astrochemistry setups.

The CO ice band strength provided in this work, 
$\mathcal{A}(CO)$ = (8.7 $\pm$ 0.5) $\times$ 10$^{-18}$ cm ~molecule$^{-1}$, 
leads to column densities a factor of about 1.3 larger than the most
frequently used literature value from \cite{Jiang1975}, 
$\mathcal{A}(CO)$ = 1.1 $\times$ 10$^{-17}$ cm
~molecule$^{-1}$. The updated CO band strength in \cite{Bouilloud_2015} considered a
refractive index of $n$ = 1.25 and density of $\rho$ = 0.8 g
cm$^{-3}$, it is very similar to the value of \cite{Jiang1975}. As explained in 
Sect. \ref{sect:band_strength_estimation}, our lower value of
$\mathcal{A}(CO)$ was obtained from our measurements of ice thickness
and IR absorbance, and the values reported in \cite{LUNA2022_aa} for the
CO ice refractive index and density measured at 20 K deposition temperature, respectively 
$n$ = 1.30  and $\rho$ = 0.876 g cm$^{-3}$.          

\section{Astrophysical implications} \label{astro_implications}

The CO ice density values obtained in this work are in line with those
recently reported by \cite{LUNA2022_aa}. Moreover, we observed a decrease of the density at deposition temperatures below 13 K. This behaviour of the density reminds that of color in analog CO ice experiments, where the
color temperature increases in the same deposition temperature range. Eye
observations see a brownish color at 8 K that becomes gradually less
intense up to 13 K. At higher deposition temperatures, the CO ice is
translucent and becomes almost transparent when the temperature
approaches the thermal desorption of the ice, suggesting that the ice is
nearly crystalline \citep{Carrascosa_2021}. These colorimetric
measurements and the linear drop of the CO photodesorption rate \citep{Oberg2007,Oberg2009, MunozCaro2016, Sie2022} for increasing
deposition temperature, might be a manifestation of molecular disorder
in CO ice grown below 20 K. If this is correct, CO ice grown at the
lowest investigated temperature, around 10 K, presents the highest
molecular disorder, and we report here that this CO ice structure corresponds to the lowest ice density.                

\cite{Jiang1975} estimated the infrared band strength of CO ice deposited at high pressure compared to modern setups: their deposition rate of 0.5 to 2 $\mu$m requires a pressure about 16 to 66 times higher than typical experiments performed at 1 $\times$ 10$^{-6}$ mbar during deposition. To obtain this band strength, \cite{Jiang1975} also adopted a literature value of the CO ice density measured at a relatively high temperature, 30 K \citep{Vegard1930}. According to \cite{LUNA2022_aa} and this work, the CO ice density depends on the temperature and pressure during deposition, and therefore, the value of the CO infrared band strength in \cite{Jiang1975} needs a revision. 
We propose to use 
$\mathcal{A}(CO)$ = (8.7 $\pm$ 0.5) $\times$ 10$^{-18}$ cm ~molecule$^{-1}$ for future column density
estimations. This value is valid in the experimental range from 11 to 28 K deposition temperatures investigated in this work, for which no variations of the integrated absorbance area and sticking probability were found \citep{Cazaux_2017}. The CO ice column density values reported in previous experimental and observational
papers might thus be underestimated, they would be about 23 \% lower than
the actual value. Most of the CO ice column densities reported in the
literature adopted a band strength of 
$\mathcal{A}(CO)$ = 1.1 $\times$ 10$^{-17}$ cm ~molecule$^{-1}$ \citep{Jiang1975}, they would need to be multiplied by a factor of 1.3 for correction. An example is the number of monolayers on the surface, or just beneath the surface of the ice, involved in the photodesorption of CO, i.e. $N$ = 5 $\times$ 10$^{15}$ molecules cm$^{-2}$ or about 5 monolayers (ML) where 1 ML = 1 $\times$ 10$^{15}$ molecules cm$^{-2}$ \citep{MunozCaro2010, Fayolle2011, Chen_2014}. After correction, this becomes 
6.5 ML. Considering our average density value of CO ice deposited at 11 K, 0.837 g cm$^{-3}$, and Eq. \ref{eq:dens}, it is found that the thickness of one monolayer is 0.56 nm, and 3.36 nm for the top 6.5 ML of ice. The UV photons emitted by the MDHL that are absorbed deeper than 3.36 nm would not lead to photodesorption of CO molecules.   

\section*{Acknowledgements}
\label{acknowledgements}
This research has been funded by projects PID2020-118974GB-C21, PID2020-118974GB-C22, and  MDM-2017-0737 Unidad de Excelencia ‘‘Mar\'{i}a de Maeztu’’-- Centro de Astrobiolog\'{i}a (INTA-CSIC) by the Spanish Ministry of Science and Innovation, and grant No. NSTC 110-2628-008-004-MY4 from Taiwan.
The student G. Mettepenningen from TU Delft participated in the preliminary phase of this project. 

\section*{Data Availability}
The data underlying this article cannot be shared publicly. 


\bibliographystyle{mnras}
\bibliography{GonzalezDiaz_mnras_Laser} 

\bsp	
\label{lastpage}
\end{document}